\documentclass[aps,pra,floatfix,superscriptaddress,twocolumn,showpacs]{revtex4}
\usepackage{amsfonts}
\usepackage{amssymb}
\usepackage{amsmath}
\usepackage{graphicx,psfrag}
\usepackage{bm}
\usepackage[T1]{fontenc}
\usepackage{calligra}
\usepackage{dcolumn}    
\usepackage{epsf}  
\usepackage{epsfig}      
\usepackage{times}  
\usepackage{nicefrac}  
\usepackage{color}      
  
\newcolumntype{.}{D{x}{}{-1}}  

\begin{document}

\title{Effect of polydispersity to specific absorption rate}

\author{Zs. J\'{a}nosfalvi$^1$\footnote{corresponding author: janosfalvi.zsuzsa@atomki.mta.hu} , J. Hakl$^1$ and K. Vad}
\affiliation{Institute of Nuclear Research, P.O.Box 51, H-4001 Debrecen, Hungary}

\begin{abstract} 

The predictions of a newly developed Bloch-Bloembergen  alike analytic magnetization model are compared to experimental results. The effect of size polydipersity on  the specific absorption loss is demonstrated for the magnetic nanoparticles containing media. Specific absorption rate shows resonance like behivior as a function of particle size. The obtained results are in excellent agreement with experimental data. The dominace of the N\'{e}el relaxation over the Brownian one is demonstrated.

\end{abstract}


\maketitle 

\section{Introduction}
\label{sec_intro}

Magnetic fluid hyperthermia (MFH) has gained wide interest in its applicability in medical sciences \cite{biomedical}. Enhancing specific absorption loss (SAR) by core-shell nanostructures, see e.g. \cite{boubeta,lee} or via anisotropy \cite{bertotti, poperechny} are also extensively studied. Furthermore, efficiency of circularly polarized field opposed to linearly polarized field is also investigated by models \cite{mazsipeter,cantillon-murphy,mazsijudit, raikher,denisov2006prb,sun,denisov2006prl} and experiments \cite{ahsen, jordan} as well. An interesting study of the chain formation of nanoparticles can be read in the paper of He \cite{he}, examining the possibility of data storage.

In MFH, the basic question is how to describe energy losses. In the literature relaxation and hysteresis losses are distinguished though both proportional to the area of hysteresis curve. See also the comment about it in the paper of Carrey \textit{et al.} \cite{carrey}, in which they also argue the necessity of distinction. The relaxation losses could originate either turning the single-domain particle with its magnetic momentum, called Brownian relaxation, or the nanoparticle itself is fixed but its magnetic momentum aligns to the external field, as in the case of N\'{e}el relaxation. There is a vivid dispute, whether both relaxations contribute to the losses and if so, to what extent. Also in the paper of Wang \textit{et al.}\cite{wang}, they compared SAR values for nanoparticles with and without polymerization, and they found no change in the SAR value excluding the possibility of Brownian relaxation. Furthermore from theoretical considerations, at typical size of nanoparticles up to few tens nanometers of diameters, depending on the experimental conditions, N\'{e}el relaxation is regarded to be the dominant process. 

It is also a question, whether a single independent particle or statistical ensemble of particles should be taken into account. The single particle picture has the advantage of ab initio description, though in this case no temperature and therefore thermal fluctuation is included yet \cite{mazsipeter,mazsijudit}, while in the statistical picture the treatment is phenomenological \cite{shliomis,rosensweig1985}. Shliomis \cite{shliomis} provided a complex equation of motion of the magnetization from hydrodynamical considerations, but as we saw till now, radical simplifications were used leaving only the Debye-term of the equation, see e.g. paper of Cantillon-Murphy \textit{et al}. \cite{cantillon-murphy}. Stochasticity can also be included \cite{raikher} for a more elaborated picture of possible relaxation processes. 

Aim of this paper is to test the validity of magnetization dynamics model with Larmor-precession term included  \cite{jzsuzsa} in comparison with experimental results \cite{mehdaoui,wang,hergt,suto}.  In the model the so-called Bloch-Bloembergen equation\cite{bloch, bloembergen} is rewritten for the single domain nanoparticle magnetization. We show that experimental results can be understood by the polidispersity of size of nanoparticles. Doe to analytic solution  for both linearly and circularly polarized fields, the model is unique being valid without restrictions to any parameter value. 

For example, in the paper of Mehdaoui \textit{et al.}\cite{mehdaoui}, a combined theoretical and experimental study was conducted, in which linear response theory and Stoner-Wohlfarth model were used, according to their range of validity.  We provided reliable SAR values with our model both in the range of linear response theory and Stoner-Wohlfarth model. 

For practical reason, we focus on hyperthermia application and carried out our analysis and evaluation at lower field strength and frequency region. In this case we can claim the equivalence of linearly and circularly polarized field for SAR. Finally, dominance of N\'{e}el relaxation with respect to Brownian relaxation is also discussed. 

\section{Theoretical background}
\label{sec_theo}

 
\subsection{Basic equations of motion}
\label{subsec_eqs}
The basic equation of motion of the magnetization is given as,

\begin{equation}
	d\bm{M}/dt = \gamma\bm{M} \times \bm{B}.
	\label{dynamicsM}
\end{equation}

Here $\gamma$ is the gyromagnetic ratio with value $\gamma= - 1.76 \times 10^{11}$ Am$^2$/Js. It comes from eq.(\ref{dynamicsM}) that any change of the magnetization is perpendicular to $\bm{M}$ and for constant $\bm{B}$ the angle between the two vectors $\bm{M}$ and $\bm{B}$ is also constant. The solution is the precessing $\bm{M}$ around $\bm{B}$ for this equation of motion, yielding the \textit{Larmor precession}, where $\bm{\omega_L} = \gamma\bm{M}\times\bm{B}/M_{\bot}$. Here $M_{\bot}$ is the projection of $\bm{M}$ on the plane perpendicular to $\bm{B}$. The Larmor frequency is defined as a positive quantity, $\omega_L=|\gamma| B$. 

Shliomis in $1974$ \cite{shliomis} has suggested the equation of exponential relaxation to be written for describing the behaviour of magnetic nanoparticles in ferrofluids at given conditions,

\begin{equation}
\frac{d\bm{M}}{dt}=-\frac{\bm{M}-\bm{M}_{eq}}{\tau},
	\label{debye}
\end{equation}

\noindent where $\bm{M}$ is the average magnetization of the particles, 

\begin{equation}
\bm{M}_{eq}=M_S {\cal L}\left( \frac{\mu_0 HM_d V}{kT} \right)\hat{\bm{e}}_H,
	\label{meq}
\end{equation}

\noindent $V$ is the particle volume, $\hat{\bm{e}}_H=\bm{H}/H$ is the unit vector along $\bm{H}$ and $M_S$ is the saturation magnetization of the colloid $M_S=\phi M_d$, with volume fraction $\phi$ and the single-domain magnetic nanoparticle magnetization $M_d$. The Langevin function, ${\cal L}(x)=\coth(x)-1/x$, gives the magnitude of the magnetization in thermal equilibrium, where $M_{eq}$ must be an ensemble average and so is $\bm{M}$. The eq.(\ref{debye}) is the reduced form of the Shliomis relaxation equation \cite{shliomis} and often called Debye relaxation equation.  Susceptibility is frequently approximated with Rosensweig's chord susceptibility \cite{rosensweig2002},

\begin{equation}
\chi_{ch}=\frac{M_S}{H_0} {\cal L}\left( \frac{\mu_0 H_0 M_d V}{kT} \right)
	\label{chord}
\end{equation}

\noindent instead of the Langevin function appearing in eq.(\ref{meq}), where actual magnetic field is used contrary to its maximum.

The Debye relaxation equation, eq.(\ref{debye}), can be regarded as a simplified version of the Bloch-Bloembergen equation \cite{bloch, bloembergen}, which describes nuclear magnetic resonance experiments  \cite{bloch} and ferromagnetic resonance \cite{bloembergen}. The latter does include the gyromagnetic torque, $\mu_0 \gamma \bm{M}\times\bm{H}$. The Bloch-Bloembergen alike formalism was used in paper \cite{jzsuzsa} for the description of the dynamics of magnetization of paramagnetic nanoparticles in ferrofluids. It is pointed out that at field strengths and frequencies used in MFH, the linearly and circularly polarized fields can be equivalent up to a normalization constant, which holds for the interested hyperthermia region in this paper.


\subsection{Model equations}
\label{subsec_bb}

Analytical solutions of Bloch-Bloembergen alike equations are briefly summarized here for circularly and linearly polarized fields:

\begin{equation}
\frac{d\bm{M}}{dt}=\mu_0\gamma \bm{M}\times\bm{H} -\frac{\bm{M}-\bm{M}_{eq}}{\tau}, 
	\label{debyewlarmor}
\end{equation}

\noindent with $\bm{M}_{eq}$ as given in eq.(\ref{meq}). For details of the derivation of equations, see paper \cite{jzsuzsa}. 

Setting rotating magnetic field, $H_x=H_0 \cos(\omega t); H_y=H_0 \sin(\omega t)$, the solution is:

\begin{equation}
\bm{M}(t)=M_{eq}(H_0)\frac{\omega}{\Omega^2}\frac{1}{1+(\Omega\tau)^2}
\begin{pmatrix}
\omega\cos(\omega t)+\Omega^2\tau\sin(\omega t)\\ \omega\sin(\omega t)-\Omega^2\tau\cos(\omega t)\\ -\omega_L
\end{pmatrix},
	\label{analfinal}
\end{equation}

\noindent where $\Omega = \sqrt{\omega^2+\omega_L^2}$ and $\omega_L=\mu_0|\gamma|H_0$. Calculating the energy loss per cycle,

\begin{eqnarray}
E&=&-\mu_0 \int_0^{0+2\pi/\omega} \bm{M}\cdot\frac{d\bm{H}}{dt}\ dt\nonumber\\
&=&2\pi\mu_0 M_{eq}(H_0)H_0 \frac{\omega}{\Omega}\frac{\Omega\tau}{1+(\Omega\tau)^2}.
	\label{energyCirc}
\end{eqnarray}

\noindent The specific absorption rate (SAR) is defined as energy loss for unit time and mass:

\begin{equation}
SAR \doteq \frac{E\omega}{2\pi\rho}=\frac{\mu_0 M_{eq}(H_0)H_0}{\rho}\frac{\omega^2}{\Omega}\frac{\Omega\tau}{1+(\Omega\tau)^2}.
\label{SARCirc}
\end{equation}

In case of linearly polarized field, $\bm{H}(t)=(0,0,H_0\cos(\omega t))$, the solution for $M_z(t)$ is found in the form of effective magnetization:

\begin{equation}
M_z^{eff}(t)=2M_S\sum_{m=1}^{\infty}\frac{1}{(m!)^2}B_{2m}\zeta^{2m-1}\frac{\cos(\omega t)+\omega\tau\sin(\omega t)}{1+(\omega\tau)^2},
	\label{effsolz}
\end{equation}

\noindent where $B_{2m}$ and $\zeta^{2m-1}$ are the Bernoulli numbers [Ref.~\onlinecite{gradshteyn}, p.1040 and p.1045] and the argument of the Langevin function $\zeta \doteq \mu_0 H_0 M_d V/kT$ respectively. 

The energy loss per cycle is

\begin{widetext}
\begin{equation}
E=-\mu_0 \int^{t_0+2\pi/\omega}_{t_0} \bm{M}\cdot\frac{d\bm{H}}{dt}dt=-\mu_0 \int^{t_0+2\pi/\omega}_{t_0} M_z^{eff}\cdot\frac{dH_z}{dt}dt=\pi\mu_0\chi_{loss}(\zeta)H_0\frac{\omega\tau}{1+(\omega\tau)^2},
	\label{energyLin}
\end{equation}
\end{widetext}

and the corresponding SAR is 

\begin{equation}
SAR \doteq \frac{E\omega}{2\pi\rho}=\frac{\mu_0  \chi_{loss}(\zeta)H_0}{2\rho}\omega\frac{\omega\tau}{1+(\omega\tau)^2}.
\label{SARLin}
\end{equation}


\section{Comparisons with experimental results and discussion}
\label{sec_results}

In this chapter our model predictions are compered to experimental data.  All the parameters can be found in more detail in references. First of all, we highlight the paper of Mehdaoui \textit{et al.} \cite{mehdaoui}, because they included very detailed information on polydispersity of their samples and thorough analysis of their measurement. It provided great opportunity to test our model with polydispersity built in. Excellent agreement was found and it encouraged us to analize other resutls when polydispersity is mentioned but no additional data are given. Log-normal distribution was used like in the paper of Mehdaoui \textit{et al.} \cite{mehdaoui}, as particle size could take only non-negative values. Fitting log-normal distribution to experimental data, we can conclude not only the possible standard deviations from the average value but also might distinguish among samples fabricated by different processes. 

Analyzing samples from Mehdaoui \textit{et al.}, see Table \ref{tab:mehdaoui}, we found excellent agreement both for SAR and standard deviation reproducing also the particle size distribution with our model. Mehdaoui \textit{et al.} used samples with mean radius $2.8$nm, $9.85$nm and $13.75$nm and found ferromagnetic behavior with $r\ge 7$nm. Table \ref{tab:mehdaoui} shows an other very important feature of our model providing reliable SAR values out of the range of linear response theory, where originally Stoner-Wohlfarth model was used instead. Accordingly we carried out analytical simulation for direct comparisons with experiments instead of numerical ones.  Finally, we included polydispersity in our model.

\begin{table}
	\centering
		\begin{tabular}{|c|c|c|c|}
		
		\hline
		 & $\mathbf{r= 2.8nm}$ & $\mathbf{r= 9.85nm}$ & $\mathbf{r= 13.75nm}$ \\
		\hline
		$SAR_{Mehdaoui} (W/g)$ & 20 & 290 & 40 \\
		\hline
		$SAR_{N} (W/g)$ & 21 & 291 & 42.5 \\ 
		\hline
		$\sigma_{Mehdaoui}$ & 0.2 & 0.09 & 0.2 \\ 
		\hline
		$\sigma_{calc}$ & 0.16 & 0.13 & 0.19 \\
		\hline
		$\pm\delta_{calc}(nm)$ & -1.2,\ +1.8 & -3.2, +5.3 & -6.3, 11.8 \\
		\hline
			
		\end{tabular}
	\caption{SAR (W/g) values are compared for different nanoparticles sizes. SAR from paper Mehdaoui \textit{et al.} \cite{mehdaoui} ($SAR_{Mehdaoui}$) and SAR calculated taking N\'{e}el-relaxation time ($SAR_{N}$) only are shown. Log-normal distribution is taken. Their standard deviations are also given by Mehdaoui \textit{et al.} ($\sigma_{Mehdaoui}$) and by us ($\sigma_{calc}$). The possible deviations from the mean value ($\pm\delta_{calc}$) are given in the last row. }
	\label{tab:mehdaoui}
\end{table}

In Table \ref{tab:nopoly} experimental SAR data from Wang \textit{et al.} \cite{wang}, Hergt \textit{et al.} \cite{hergt} and Suto \textit{et al.} \cite{suto} are compared to model predictions calculated without polydispersity. It can be seen, the predictions are at the order of experiments, but inconsistently under- or overestimating them. Data also show that taking into account only Brownian relaxation results in extremely poor agreement among model predictions and experiments. Including polydispersity in calculations results in drastic improvement in the consistency of experimental data and model prediction, the error is less than $5\%$ in most cases. In case of pure Brownian relaxation process the agreement is poor again. Based on these results, the scenario of pure  Brownian relaxation process can be excluded. Whether Brownian relaxation process is still contributing to the relaxation processes or only N\'{e}el relaxation should be considered, cannot be decided. The  first possibility slightly overestimates, the latter one underestimates the experiments.

\begin{table}
	\centering
		\begin{tabular}{|c|c|c|c|}
		
		\hline
		 & \textbf{Wang \textit{et al.}} & \textbf{Hergt \textit{et al.}} & \textbf{Suto \textit{et al.}} \\
		\hline
		$SAR_{Paper} (W/g)$ & 123 & 45 & 22 \\
		\hline
		$SAR_{N} (W/g)$ & 3 & 19 & 30 \\ 
		\hline
		$SAR_{tot} (W/g)$ & 3 & 19 & 35 \\
		\hline
		$SAR_{B} (W/g)$ & 767 & 1313* & 5.5 \\
		\hline
		$r (nm)$ & 2.5 & 2.8 & 3.125 \\
		\hline
			
		\end{tabular}
	\caption{Without polydispersity, SAR (W/g) values are compared for different papers. SAR from Wang \textit{et al.} \cite{wang}, Hergt \textit{et al.} \cite{hergt} and Suto \textit{et al.} \cite{suto} are listed. Samples were for Wang \textit{et al.} size of $r=5nm$ and for Suto \textit{et al.} the so-called $Sample A$. SAR calculated taking N\'{e}el-relaxation time ($SAR_{N}$) only, combined N\'{e}el- and Brownian-relaxation time ($SAR_{tot}$) and Brownian-relaxation time ($SAR_{B}$) only.
	* No assumption for the value of hydrostatic volume is taken. }
	\label{tab:nopoly}
\end{table}

\begin{table}
	\centering
		\begin{tabular}{|c|c|c|c|}
		
		\hline
		 & \textbf{Wang \textit{et al.}} & \textbf{Hergt \textit{et al.}} & \textbf{Suto \textit{et al.}} \\
		\hline
		$SAR_{Paper} (W/g)$ & 123 & 45 & 22 \\
		\hline
		$SAR_{N} (W/g)$ & 122 & 47 & 19 \\ 
		\hline
		$SAR_{tot} (W/g)$ & 129 & 47 & 23 \\
		\hline
		$SAR_{B} (W/g)$ & 146 & 146* & 5.8 \\
		\hline
		$r (nm)$ & 2.5 & 2.8 & 3.125 \\
		\hline
		$\sigma_{calc}$ & 0.27 & 0.14 & 0.25 \\ 
		\hline
		$\pm\delta_{calc}(nm)$ & -1.5,\ +6.5 & -1.0, +3.0 & -2.0, +5.5 \\
		\hline
			
		\end{tabular}
	\caption{With polydispersity, SAR (W/g) values are compared for different papers. SAR from Wang \textit{et al.} \cite{wang}, Hergt \textit{et al.} \cite{hergt} and Suto \textit{et al.} \cite{suto} are listed. Samples were for Wang \textit{et al.} size of $r=5nm$ and for Suto \textit{et al.} the so-called $Sample A$. SAR calculated taking N\'{e}el-relaxation time ($SAR_{N}$) only, combined N\'{e}el- and Brownian-relaxation time ($SAR_{tot}$) and Brownian-relaxation time ($SAR_{B}$) only. Their standard deviations ($\sigma_{calc}$) are calculated by us only. The possible deviations from the mean value ($\pm\delta_{calc}$) are given in the last row.
	* No assumption for the value of hydrostatic volume is taken. }
	\label{tab:withpoly}
\end{table}

\begin{table}
	\centering
		\begin{tabular}{|c|c|c|c|}
		
		\hline
		 & $\mathbf{r= 5nm}$ & $\mathbf{r= 4nm}$ & $\mathbf{r= 3nm}$ \\
		\hline
		$SAR_{Wang} (W/g)$ & 123 & 78 & 50 \\
		\hline
		$SAR_{N} (W/g)$ & 122 & 76 & 47 \\ 
		\hline
		$SAR_{tot} (W/g)$ & 129 & 82 & 53 \\ 
		\hline
		$SAR_{B} (W/g)$ & 146 & 496 & 304 \\ 
		\hline
		$\sigma_{calc}$ & 0.27 & 0.36 & 0.47 \\ 
		\hline
		$\pm\delta_{calc}(nm)$ & -2.6,\ +7.4 & -2.0, +7.5 & -2.0, +8.5 \\
		\hline
			
		\end{tabular}
	\caption{SAR (W/g) values are compared for different nanoparticles sizes. SAR from paper Wang \textit{et al.} \cite{wang} ($SAR_{Wang}$) and SAR calculated taking N\'{e}el-relaxation time ($SAR_{N}$) only, combined N\'{e}el- and Brownian-relaxation time ($SAR_{tot}$) and Brownian-relaxation time ($SAR_{B}$) only are shown. Log-normal distribution is taken. Their standard deviations ($\sigma_{calc}$) are calculated by us only. The possible deviations from the mean value ($\pm\delta_{calc}$) are given in the last row. }
	\label{tab:wang}
\end{table}

Finally, we  compared our model predictions to a series of data taken from Wang \textit{et al.} \cite{wang}. We experienced the same tendencies as before. The consideration of Brownian relaxation process exclusively did not explain the experimental data. Contrary, taking N\'{e}el relaxation alone, resulted in slight underestimation of measured values,  the error was within $5\%$. It might worth be note, the absolute value of variation of the particle size from the mean value is nearly constant, which could be explained by having samples fabricated with the same technique. Contrary the samples of  Mehdaoui \textit{et al.} were prepared differently, and according to it, their standard deviations of particle sizes differ markedly. The asymmetry of the variation of particle size from the mean value can be explained by the asymmetry of the log-normal distribution. 

\begin{figure}
	\centering
		\includegraphics{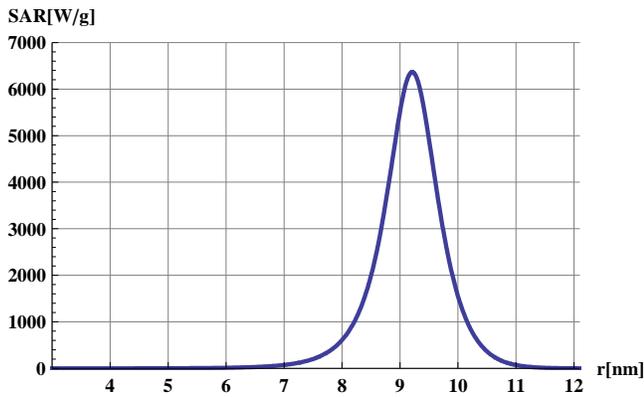}
	\caption{SAR is shown as a function of particle radius. Resonance can be observed at $9.21$ nm with $SAR=6365$ W/g. }
	\label{fig:janosfalvi_Fig1}
\end{figure}

\begin{figure}[b]
	\centering
		\includegraphics{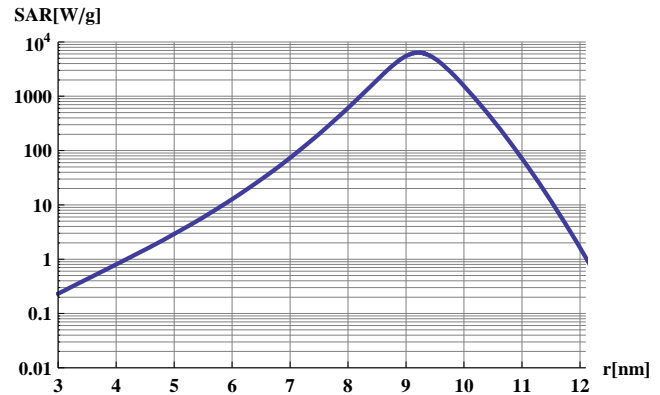}
	\caption{SAR is shown as a function of particle radius at logarithmic scale. Low SAR data are enhanced by logarithmic scale. Resonance can be observed at $9.21$ nm with $SAR=6365$ W/g.}
	\label{fig:janosfalvi_Fig2}
\end{figure}

Wang \textit{et al.} also estimated the optimum size for reaching the maximum of SAR being at $r=9.15$nm, based on the linear response theory and Rosensweig's susceptibility. From our model, we got $r=9.21$nm for the optimum size to get resonance, which size is in excellent agreement with the estimations given by Wang \textit{et al.}. We note, that at the resonance the SAR is $6365 W/g$, see Fig.\ref{fig:janosfalvi_Fig1}. The same curve can be seen on logarithmic scale, Fig.\ref{fig:janosfalvi_Fig2}, to highlight the tendency for small nanoparticles. 

\section{Summary}
\label{sec_sum}

The aim of this paper was to show the effect of size polydispersity on the specific absorption loss in a magnetic nanoparticles containing media. For the first time, analytic calculations were carried out instead of numerical simulations, using a Bloch-Bloembergen alike model, developed and described detailed in paper \cite{jzsuzsa}. As a reference, the paper of Mehdaoui \textit{et al.} was used for direct comparison, including nanoparticles $r\ge 7$nm, the range of Stoner-Wohlfarth model beyond linear response theory. We found excellent agreement, see Table\ref{tab:mehdaoui}. Based on this agreement, we estimated the possible mean value and standard deviation of particle sizes for other experiments of Wang \textit{et al.}, Hergt \textit{et al.} and Suto \textit{et al.}, see Table\ref{tab:withpoly}. For the sake of comparison, the SAR values were calculated also without polydispersity, Table\ref{tab:nopoly}, and we found drastic improvement when polydispersity was included. Our analytical approach predicted the resonance in the experiment of Wang \textit{et al.}. After analyzing standard deviations from mean value of particle sizes, samples fabricated by different technologies might be distinguished, see experiments of Mehdaoui \textit{et al.} and Wang \textit{et al.} It can be concluded, the polydispersity is inevitable in experiments and it should be taken into account in models for calculating SAR as well. The case of pure Brownian relaxation is excluded. N\'{e}el relaxation was found to be essential in relaxation processes for paramagnetic nanoparticles at hyperthermia. 

\section*{Acknowledgement}
One of the author, Zs.J., wishes to express special thanks to I. N\'andori, P.F. de Ch\^{a}tel and to colleagues in the lab for their valuable critical advices and continuous support. The authors acknowledge support from the Hungarian Scientific Research Fund (OTKA) No.101329.


\begin{thebibliography}{99}


\bibitem{biomedical}
Q. A. Pankhurst, J. Connolly, S. K. Jones, and J. Dobson,\textit{ J. Phys. D: Appl. Phys.} \textbf{2003}, \textit{36}, R167;
M. Ferrari, \textit{Nat. Rev. Cancer} \textbf{2005}, \textit{5}, 161;
R. Hergt, S. Dutz, R. Muller, M. Zeisberger, \textit{J. Phys.: Condens. Matter} \textbf{2006}, \textit{18}, S2919;
R. Hergt, S. Dutz, \textit{J. Magn. Magn. Mater.} \textbf{2007}, \textit{311}, 187.
S. Laurent, D. Forge, M. Port, A. Roch, C. Robic, L. Vander Elst, and R. N. Muller, \textit{Chem. Rev.} \textbf{2008}, \textit{108}, 2064;
J. D. Alper, \textit{Thesis (Ph.D.)} \textbf{2010}, MIT; 
Quian Wang and Jing Liu, \textit{Fundamental Biomedical Technologies} \textbf{2011}, \textit{5}, 567-598, 
Spinger Science+Business Media: {\em Intercellular Delivery} (ed. A. Prokop);
A. L. E. Rast, \textit{Thesis (Ph.D.)} \textbf{2011}, University of Alabama, Birmingham;
D. E. Bordelon, {\em et al.}, \textit{J. Appl. Phys.} \textbf{2011}, \textit{109}, 124904;
A Arakaki, {\em et al.}, \textit{Polymer Journal} \textbf{2012}, doi:10.1038/pj.2012.32.

\bibitem{boubeta} Carlos Martinez-Boubeta, Konstantinos Simeonidis, David Serantes, Iv\'{a}n Conde-Lebor\'{a}n, Ioannis Kazakis, George Stefanou, Luis Pe\~{n}a, Regina Galceran, Lluis Balcells, Claude Monty, Daniel Baldomir, Manassis Mitrakas and Makis Angelakeris, 
Adjustable Hyperthermia Response of Self-Assembled Ferromagnetic Fe-MgO Core-Shell Nanoparticles by Tuning Dipole-Dipole Interaction, 
\textit{Advanced Functional Materials} \textbf{2012}, \textit{22}, 3737-3744

\bibitem{lee} Jae-Hyun Lee, Jung-tak Jang, Jin-sil Choi, Seung Ho Moon, Seung-hyun Noh, Ji-wook Kim, Jin-Gyu Kim, Il-Sun Kim, Kook In Park and Jinwoo Cheon, 
Exchange-coupled magnetic nanoparticles for efficient heat induction, 
\textit{Nature Nanotechnology} \textbf{2011}, \textit{6}, 418-422

\bibitem{bertotti} Giorgio Bertotti, Claudio Serpico, and Isaak D. Mayergoyz, 
Nonlinear Magnetization Dynamics under Circularly Polarized Field,
\textit{Phys. Rev. Lett.} \textbf{2001}, \textit{86}, 724

\bibitem{poperechny} I. S. Poperechny, Yu. L. Raikher, and V. I. Stepanov, 
Dynamic magnetic hysteresis in single-domain particles with uniaxial anisotropy,
\textit{Phys. Rev. B} \textbf{2010}, \textit{82}, 174423

\bibitem{mazsipeter} P\'{e}ter F. de Ch\^{a}tel, Istv\'{a}n N\'{a}ndori,  J\'{o}zsef Hakl, S\'{a}ndor M\'{e}sz\'{a}ros and K\'{a}lm\'{a}n Vad, 
Magnetic particle hyperthermia: N\'{e}el relaxation in magnetic nanoparticles under circularly polarized field, \textit{J. Phys.: Condens. Matter.} \textbf{2009}, \textit{21}, 124202

\bibitem{cantillon-murphy} Padraig Cantillon-Murphy, Lawrence L. Wald, Elfar Adalsteinsson and Markus Zahn, Heating in the MRI environment due to superparamagnetic fluid suspensions in a rotating magnetic field, 
\textit{Journal of Magnetism and Magnetic Materials}, \textbf{2010}, \textit{322}, 727-733 

\bibitem{mazsijudit} Istv\'{a}n N\'{a}ndori and Judit R\'{a}cz, 
Magnetic particle hyperthermia: N\'{e}el relaxation under circularly polarized field in anisotropic nanoparticles, 
\textit{Phys. Rev. E} \textbf{2012}, \textit{86}, 061404

\bibitem{raikher} Yu. L. Raikher and V. I. Stepanov, 
Power losses in a suspension of magnetic dipoles under a rotating field,
\textit{Phys. Rev. E} \textbf{2011}, \textit{83}, 021401

\bibitem{denisov2006prb} S. I. Denisov, T. V. Lyutyy, P. H\"anggi, K. N. Trohidou, 
Dynamical and thermal effects in nanoparticle systems driven by a rotating magnetic field,
\textit{Phys. Rev. B} \textbf{2006}, \textit{74}, 104406

\bibitem{sun} Z. Z. Sun, X. R. Wang, 
Strategy to reduce minimal magnetization switching field for Stoner particles,
\textit{Phys. Rev. B} \textbf{2006}, \textit{73}, 092416

\bibitem{denisov2006prl} S. I. Denisov, T. V. Lyutyy, P. H\"anggi, 
Magnetization of Nanoparticle Systems in a Rotating Magnetic Field,
\textit{Phys. Rev. Lett.} \textbf{2006}, \textit{97}, 227202

\bibitem{ahsen} Osman O. Ahsen, Ugur Yilmaz, M. Deniz Aksoy, Gulay Ertas and Ergin Atalar, 
Heating of magnetic fluid systems driven by circularly polarized magnetic field, 
\textit{Journal of Magnetism and Magnetic Materials} \textbf{2010}, \textit{322}, 3053-3059
 
\bibitem{jordan} Andreas Jordan, P. Wust, H. F\"{a}hling, W. John, A. Hinz and R. Felix, 
Inductive heating of ferrimagnetic particles and magnetic fluids: Physical evaluation of their potential for hyperthermia, 
\textit{International Journal of Hyperthermia} \textbf{2009}, \textit{25}, 499-511

\bibitem{he} Liang-Ming He, 
Relaxation of Magnetic Nanoparticle Chain without Applied Field, 
\textit{Communications in Theoretical Physics} \textbf{2011}, \textit{55}, 537-540

\bibitem{carrey} Julian Carrey, Boubker Mehdaoui, Marc Respaud, 
Simple Models for Dynamic Hysteresis Loops Calculation: Application to Hyperthermia Optimization, 
\textit{Journal of Applied Physics} \textbf{2011}, \textit{109}, 083921

\bibitem{wang} Xuman Wang, Hongchen Gu and Zhengqiang Yang, 
The heating effect of magnetic fluids in an alternating magnetic field, 
\textit{Journal of Magnetism and Magnetic Materials}, \textbf{2005}, \textit{293}, 334-340

\bibitem{shliomis}
M.I. Shliomis, \textbf{Zh. Eksp. Teor. Fiz.} \textbf{1971}, \textit{61}, 2411 [\textit{Sov. Phys. JETP} \textbf{1972}, \textit{34}, 1291];
M.I. Shliomis,\textit{ Usp. Fiz. Nauk.} \textbf{1974}, \textit{112}, 427 [\textit{Sov. Phys. Usp.} \textbf{1974}, \textit{17}, 153];.
M.I. Shliomis, \textit{Phys. Rev. E} \textbf{2001}, \textit{64}, 060501

\bibitem{rosensweig1985} R.E. Rosensweig, 
Ferrohydrodynamics, 
\textit{Cambridge University Press}, \textbf{1985}, Cambridge

\bibitem{jzsuzsa} Zs. J\'{a}nosfalvi, J. Hakl and P. F. de Ch\^{a}tel, 
Larmor precession and Debye relaxation of single-domain magnetic nanoparticles, 
\textit{arXiv:1201.5236} [cond-mat.mes-hall] \textbf{2012}

\bibitem{mehdaoui} Boubker Mehdaoui, Anca Meffre, Julian Carrey, S\'{e}bastien Lachaize, Lise-Marie Lacroix, Michel, Gougeon, Bruno Chaudret and Marc Respaud, 
Optimal Size of Nanoparticles for Magnetic Hyperthermia: A Combined Theoretical and Experimental Study, \textit{Advanced Functional Materials} \textbf{2011}, \textit{21}, 4573-4581

\bibitem{hergt} Rudolf Hergt, Wilfried Andr\"{a}, Carl G. d'Ambly, Ingrid Hilger, Werner A. Kaiser, Uwe Richter and Hans-Georg Schmidt, 
Physical Limits of Hyperthermia Using Magnetite Fine Particles, 
\textit{IEEE Transactions on Magnetics} \textbf{1998}, \textit{34}, 3745

\bibitem{suto} Makoto Suto, Yasutake Hirota, Hiroaki Mamiya, Asaya Fujita, Ryo Kasuya, Kazuyuki Tohji and Balachandran Jeyadevan, 
Heat dissipation mechanism of magnetite nanoparticles in magnetic fluid hyperthermia, 
\textit{Journal of Magnetism and Magnetic Materials} \textbf{2009}, \textit{321}, 1493-1496


\bibitem{bloch} F. Bloch, 
Nuclear Induction,
\textit{Phys. Rev.} \textbf{1946}, \textit{70}, 460

\bibitem{bloembergen} N. Bloembergen, 
On the Ferromagnetic Resonance in Nickel and Supermalloy, 
\textit{Phys. Rev.} \textbf{1950}, \textit{78}, 572

\bibitem{rosensweig2002} Ronald E. Rosensweig, 
Heating magnetic fluid with alternating magnetic field, 
\textit{Journal of Magnetism and Magnetic Materials} \textbf{2002}, \textit{252}, 370

\bibitem{gradshteyn} I.S. Gradshteyn and I.M. Ryzhik, 
\textit{Tables of integrals, series and products}, 
\textbf{7}th edition, Academic Press \textbf{2007}, 228

\end{thebibliography}
\end{document}